# Towards Lawful ISAC in Cellular Networks

**Stefano Tomasin[1*], Annalisa Volpato[2], and Marco Centenaro[1]**

[1] Dep. Of Information Engineering, University of Padova, Italy

[2] Dep. of Public, International and EU Law, University of Padova, Italy

* Corresponding author – email: Stefano.tomasin@unipd.it

## ABSTRACT

Integrated sensing and communication (ISAC) enables the tracking and detection of passive objects, including the human body, by leveraging standard wireless communication signals. While already standardized in IEEE 802.11bf for Wi-Fi, ISAC is currently being defined by 3GPP for the upcoming generations of cellular networks. This transition scales the ISAC technology from localized, uncoordinated Wi-Fi deployments into a pervasive, centralized network deployment managed by mobile network operators. However, such nationwide coverage introduces critical user privacy challenges that must comply with stringent data protection legislative frameworks, most notably the General Data Protection Regulation in the European Union. This paper provides a comprehensive analysis of applicable norms to cellular ISAC, highlights the open technical challenges in its implementation, and explores potential mitigation strategies at both the architectural and signal-processing levels, establishing tiered data access levels for various stakeholder categories and detailing how these protocols can be lawfully integrated into cellular network architectures.

## INTRODUCTION

Wireless networks have experienced a massive performance improvement in the last two decades thanks to two main solutions, i.e., Wi-Fi and mobile networks, which evolved over several standard releases. Recently, the Institute of Electrical and Electronic Engineering (IEEE) has enhanced Wi-Fi to support, along with the traditional communication service, a new *sensing service*, i.e., the capability of estimating the range, velocity, and motion of objects. Both services rely on processing the same wireless signal, which is generated for the two purposes: while communication is achieved by modulating the transmitted signal with the intended message, sensing is achieved by estimating at the receiver the effects of the propagation environment over which the signal propagates. This dual-service technology is denoted as integrated sensing and communication (ISAC). Along the same line, the third-generation partnership project (3GPP) decided to insert ISAC in forthcoming releases, and technical details are now under discussion. ISAC paves the way for new services that typically can be clustered into (i) object detection and tracking, (ii) environmental monitoring, and (iii) motion monitoring.

While ISAC poses several technical challenges, it also entails relevant legal issues. On one hand, sensing may carry information on the position, movements, and in general the behavior of humans (as better detailed in the following) potentially affect our privacy. On the other hand, the legislators are imposing strict requirements on data handling to protect privacy, with the most relevant act being the European Union (EU)'s General Data Protection Regulation (GDPR) [1]. In this perspective, the privacy risk entailed by ISAC in Wi-Fi and mobile networks is extremely different. Indeed, while ISAC data collection is localized and uncoordinated in Wi-Fi networks, it becomes pervasive and centralized in mobile networks. Moreover, on mobile networks, the information collected with ISAC can be distributed to several stakeholders at different levels. The more extensive privacy implication then calls for a tech-policy approach in the design of the upcoming sixth generation (6G) system (6GS), in which the engineering and law communities cooperate in defining technical solutions to ripe the benefit of ISAC while ensuring compliance with regulations.

The objective of this paper is to investigate such a trade-off, in order to initially identify insights that are valid across specific wireless technologies, before focusing specifically on cellular networks to characterize its ecosystem, stakeholders, and related threats, allowing to shed light on a topic that should be of interest for the design of the next releases.

## ISAC STANDARD TRENDS

ETSI has provided a general scheme for an ISAC task, which has been extended by the authors and reported in Figure 1. The *sensing task* leverages a sensing *signal*, that is exchanged between two sensing entities within a Target Sensing Service Area (TSSA), namely a *transmitter* and a *receiver*, yielding sensing *data* which, after *processing*, may produce *results* about the objects within the TSSA, e.g., characteristics of objects per se, such as type, distance, velocity, trajectory, size, shape, material, or regarding the surrounding environment. Such a simple scheme captures both the *technical steps* needed to obtain the sensing results, as well as the entailed *processing* of the sensing information before it is delivered to the *sensing user*. We will see that all these parts are relevant in achieving the privacy target.

For the technical part of sensing acquisition, we can distinguish between three types of sensing: (i) *monostatic*, when the sensing transmitter and receiver coincide, ii) *bistatic*, when transmitter and receiver are two distinct devices, and (iii) *multistatic*, when more than one transmitter and receiver are involved. Such types again are relevant for privacy since they involve different actors receiving the sensing signal, which are the first ones responsible for handling the sensed information and passing it to the processing block and sensing result users. Moreover, the signal transmitter has also the responsibility of ensuring that the signal can be used only by receivers authorized to handle the sensing information.

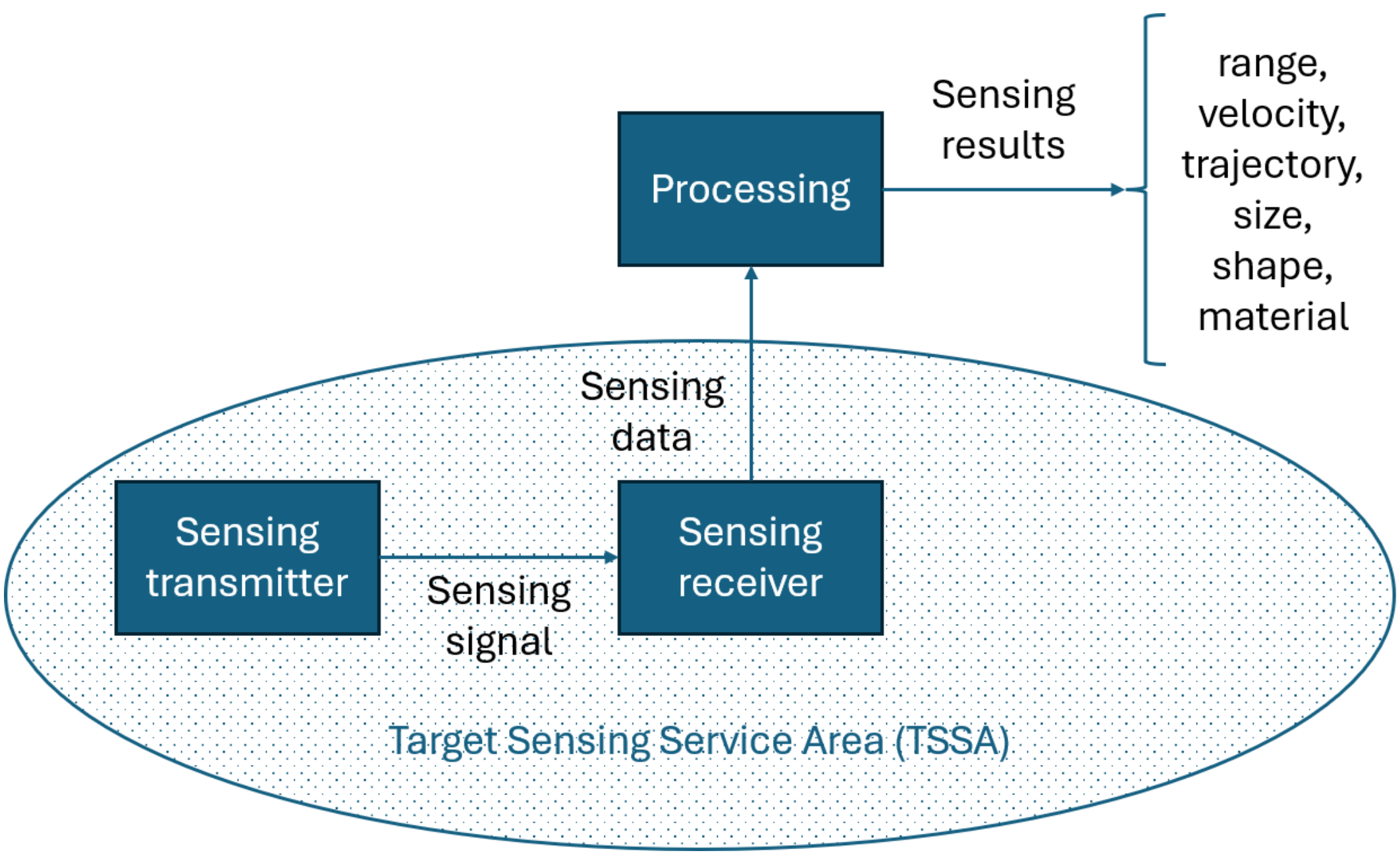

*Figure 1: Typical workflow of a sensing task.*

Notably, both prominent terrestrial wireless technologies, i.e., IEEE's Wi-Fi and 3GPP's 5G, have been progressing in integrating ISAC capabilities in their latest standard releases.

*IEEE 802.11bf.* ISAC has been introduced in the recently approved IEEE 802.11bf standard, operating both in sub-7 GHz and in the 60 GHz frequency bands. The standard primarily extracts environmental intelligence by analyzing Channel State Information (CSI) relative to orthogonal frequency-division multiplexing (OFDM) subcarriers, or, for millimeter-wave operations, by means of beamforming techniques and narrow antenna patterns. Monostatic, bistatic, and multistatic topologies are supported, with the role of sensing *initiator* and *responder* that can be taken by both access points (APs) and client stations (STAs), with sensing performance including, e.g., range resolution of 0.5-2 m, velocity resolution of 0.5 m/s, and angular resolution of 4-6 degrees.

*Privacy Requirements.* Since sensing information can be processed either at the AP or the STA, data is collected and processed locally only, even though the standard does not indicate where sensing results can be forwarded. To protect privacy, the IEEE 802.11bf standard embeds features such as MAC address randomization and secure session keys. However, these solutions only prevent unauthorized third parties from exploiting sensing traffic for illicit surveillance: the responsibility of surveillance through sensing rests on the owner of the wireless local area network.

*3GPP 5/6G.* 3GPP identified 32 sensing use cases for advanced fifth-generation (5G-Advanced) mobile networks [2], including, e.g., object and intruder detection for smart home, on a highway, for railways, for factory, for predefined secure areas around critical infrastructure, collision avoidance and trajectory tracking of unmanned aerial vehicles (UAVs), vehicles, automated guided vehicles (AGVs), public safety search and rescue, rainfall monitoring and flooding, and health and sports monitoring. Moreover, dedicated requirements on wireless sensing performance and sensing service design have already been defined for 5G-Advanced [3]. Several key performance indicators (KPIs) are defined on three general types of scenarios (object detection and tracking, environmental monitoring, and motion monitoring), including, e.g., accuracy of horizontal and vertical positioning estimate (<10m) and velocity estimate (typically <1.5m/s) as well as missed detection and false alarm probabilities (<5%). In terms of signals used for sensing, 3GPP is currently studying how to satisfy such requirements on the air interface, considering OFDM as baseline waveform [4]. Assuming that each base station can control multiple transmission-reception points (TRPs), there are 6 possible mode variants of monostatic and bistatic sensing: TRP monostatic, user equipment (UE) monostatic; TRP-TRP bistatic, TRP-UE bistatic, UE-TRP bistatic, UE-UE bistatic. Multiple types of measurements are being considered, ranging from raw data (i.e., amplitude and phase samples) per OFDM symbol at the various antenna ports of the TRPs (Level A) to object/target level measurement per base station (Level D).

An architectural description of cellular network ISAC well aligned to the general ETSI ISAC framework is provided again by ETSI through the Industry Specification Group (ISG) on ISAC. In [5], the reference model illustrated in **Errore. L'origine riferimento non è stata trovata.** has been proposed, comprising the Sensing Service *Producer* (SSP) which interacts with the Sensing Input Data *Provider* (SIDP) and the third-party Sensing Input Data Provider (3-SIDP). The former includes UEs and Access Nodes while the latter is an authorized entity, not part of 6GS, such as, e.g., an ISAC-capable Wi-Fi AP as described earlier: together, they provide input sensing data needed to implement the 6G Sensing Service. As far as the sensing procedure is concerned, the request for a new Sensing Service may originate within the 6GS by a Sensing Service *Consumer* (SSC), which may include UEs, Access Nodes, and Core Network Functions [5]. Note that a sensing service request may also reside outside the 6GS, thus denoted a third-party Sensing Service Consumer (3-SSC), to cover the case of an application function in the data network or on the UE itself requesting and consuming the 6G Sensing Service. The various entities interact according to a service-based request/response paradigm (cf. *_req and *_res interactions in Figure 2). Moreover, it is worth noting that while RAN, SSP, and SSC functions are logically part of the 6GS, their operation is not necessarily carried out *directly* by the mobile network operator (MNO). Modern network deployments increasingly rely on specialized third-party operators for managing specific network segments, making, e.g., their vendors *active* data processors rather than mere equipment suppliers [6].

When compared with the Wi-Fi scenario we note that a) the sensing information is collected across a network that may extend into an entire nation (and beyond); b) the sensing signal is received by either the gNB or the UE; c) the UE may act as a signal receiver, but then forward the sensing information to the RAN; the network includes specific functions for the processing of the sensing information and its forward to external users. Thus, in terms of privacy, ISAC in 6G is a much more challenging problem for the number and type of actors involved and the spatial extension of the sensing data collection.

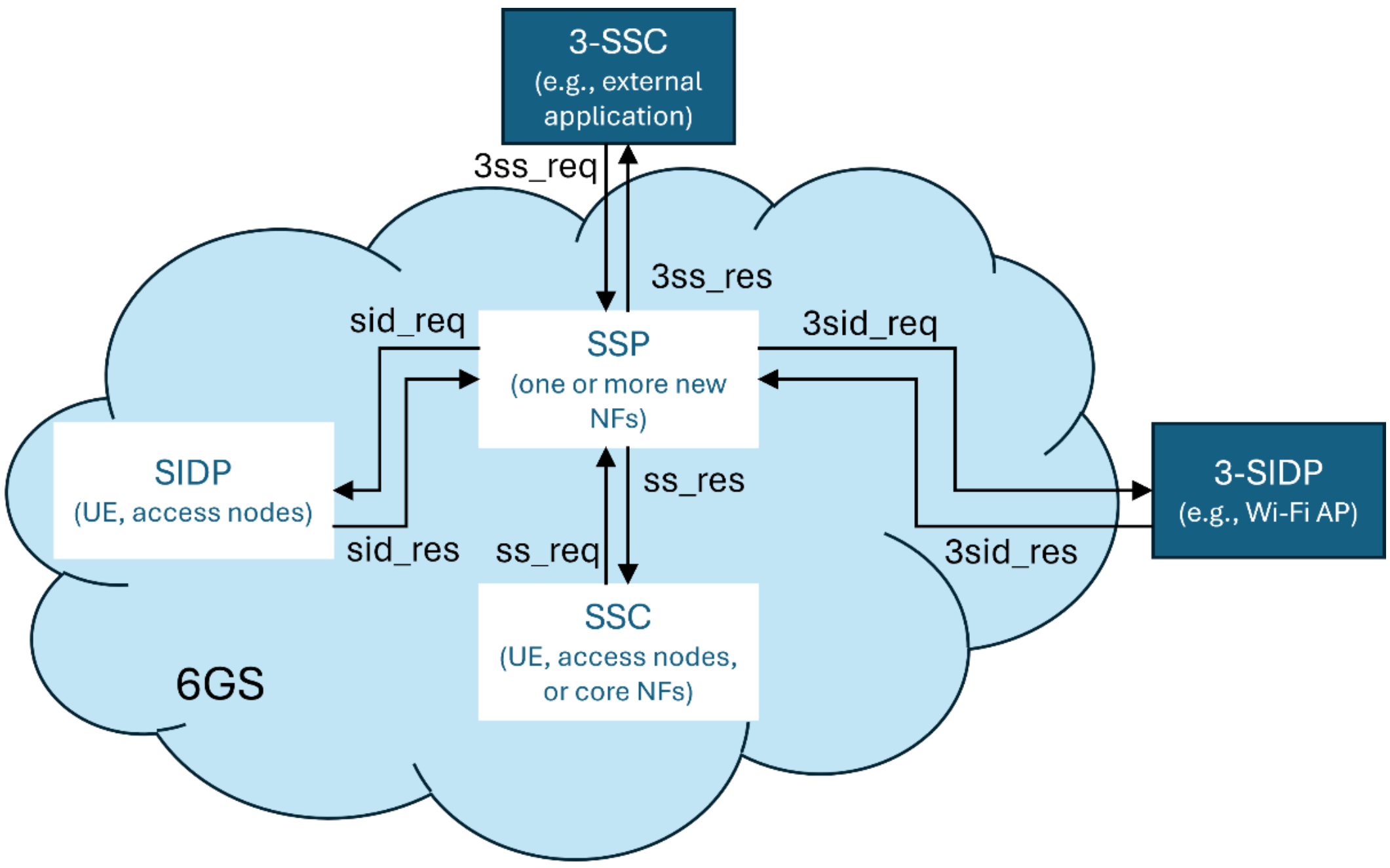

*Figure 2: 6GS reference model for the sensing service.*

*Privacy Requirements.* The necessity to address confidentiality, integrity, and privacy of sensing capabilities was already addressed in [2]. In fact, preliminary considerations dedicated to privacy are present at service requirement level in [4, Sec. 4.3, 5.2.4], however with generic statements on the need to protect the sensing data from unauthorized access, interception and eavesdropping, but also to make sure such service is following regulatory requirements. It is particularly stressed that tracking can be extended to people not carrying a UE, which makes sensing even more problematic for privacy. [5] also recognizes the need for integration of security and privacy, to which ETSI has recently devoted [7], which contains a list of key issues, description of relevant threats, and definition of potential requirement for security and privacy. Here we consider threats, introduce the concept of *signal controller* (as complementary to data controller) and thoroughly map the expended 6GS ecosystem to GDPR categorization.

Based on these considerations, we will analyze the relevant legislative framework with respect to the more impactful scenario of mobile networks.

## Relevant Legislative Framework

Privacy and data protection are fundamental rights generally protected across the world. Although GDPR specifically applies where either the processing of data is carried out by a natural or legal person established in the EU or the data concern a person who is in the EU, the GDPR sets a golden standard of legal protection which has strongly influenced the development of data protection law in many other jurisdictions, including the California Consumer Privacy Act (CCPA) in the USA [8] and Brazil's Lei Geral de Proteçäo de Dados Pessoais (LGPD) [9]. Moreover, as far as the protection of privacy in the electronic communications sector in the EU is concerned, the ePrivacy Directive [10] introduces even stronger confidentiality for the storing of information and gaining access to information stored in terminal equipment.

*Scope of Application.* The GDPR sets strict limits on the processing of any information relating to an identified or identifiable natural person (*data subject*). A person is identifiable when the information contains an identifier such as a name, an identification number, location data, an online identifier or one or more factors specific to the physical, physiological, genetic, mental, economic, cultural or social identity of that natural person. These data are considered personal data (or personal identifiable information, PII) as opposed to data anonymized or not concerning an individual, which do not fall under the scope of the GDPR. Specific data categories, such as data revealing racial or ethnic origin, political opinions, religious or philosophical beliefs, or trade union membership, genetic data, biometric data, data concerning health or a natural person's sexual life or orientation, are granted enhanced protection under Article 9 of the GDPR (sensitive data).

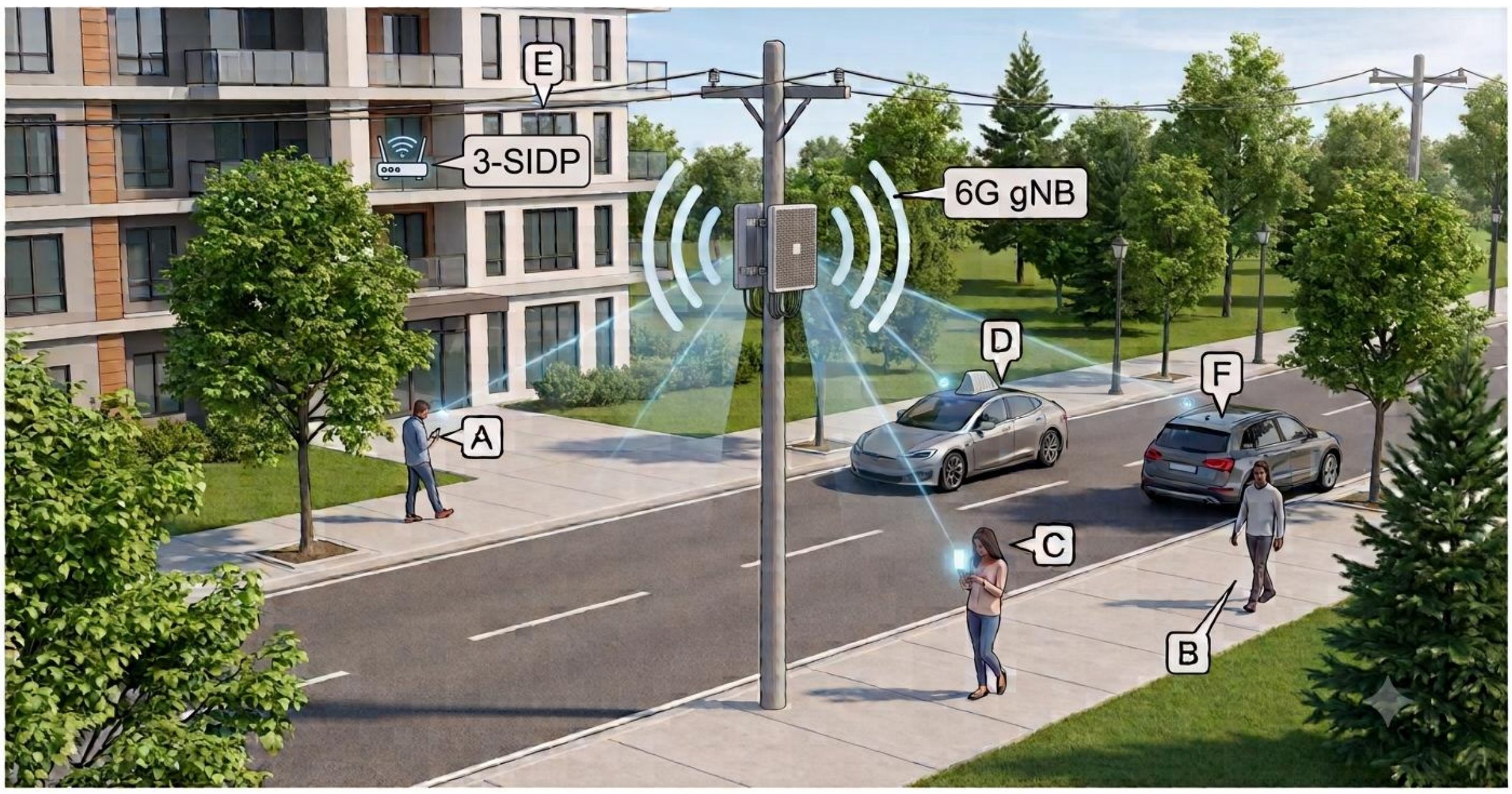

*Figure 3: Target types in 6GS.*

According to [7], there are six types of sensing targets within a TSSA, namely (see Figure 3):

- Type A: people with an ISAC-enabled device;
- Type B: people without a device;
- Type C: people with a device connected to the network for communication purposes only;
- Type D: fixed or mobile objects connected to the network for both communication and sensing (like, e.g., connected vehicles);
- Type E: fixed objects not connected to the network (such as, e.g., buildings);
- Type F: mobile objects not connected to the network (e.g., vehicles without network connectivity).

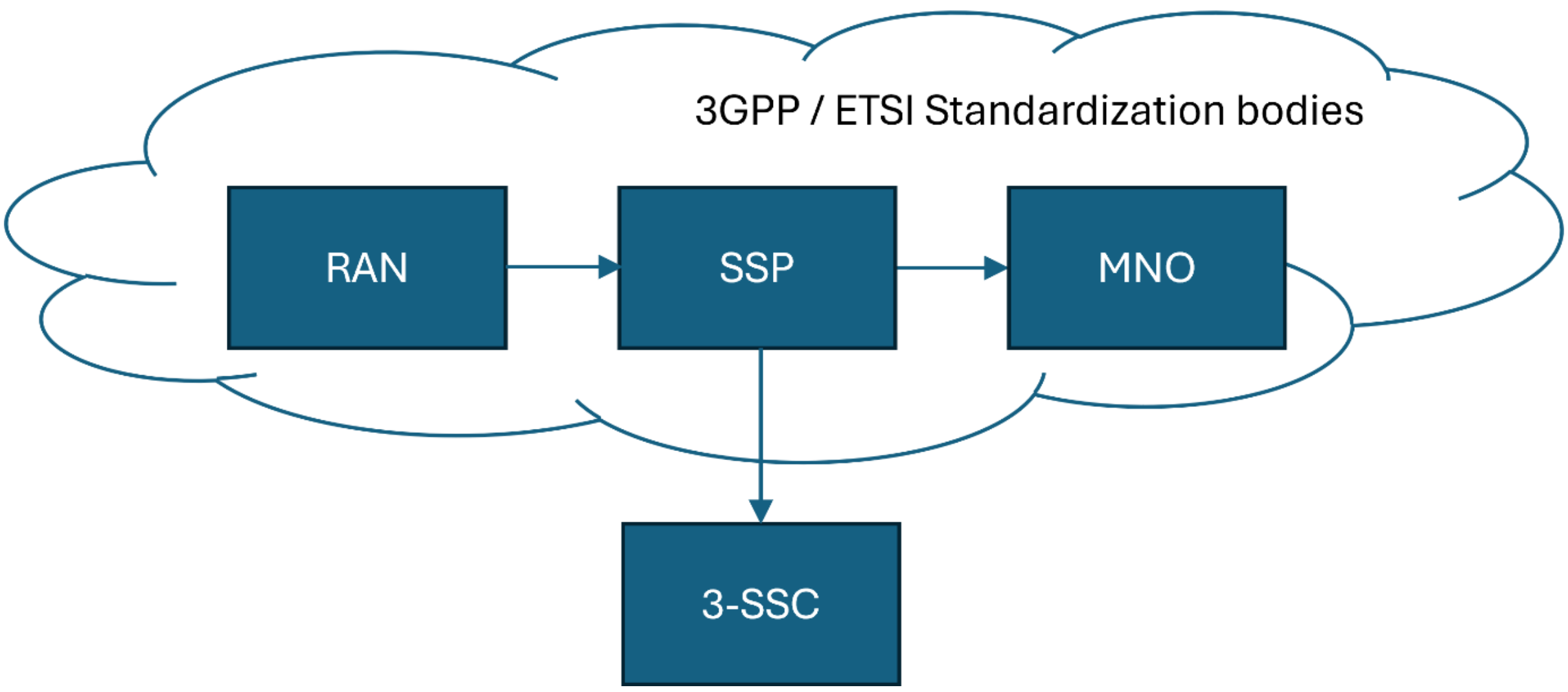

*Figure 4: Main stakeholders in the ISAC privacy of 6GS.*

Clearly, GDPR rules do not apply to Types D, E, and F categories as they do not concern humans. About the first three categories, in 6G networks the distinction between identifiable and non-identifiable person may be particularly complex due to the high amounts of data and signals, the integration with AI models, and the interaction with devices linked to specific user identifiers, which can lead to potential re-identification of anonymous data and extraction of vast amounts of personal data. In fact, while, on the one hand, ISAC-based monitoring of the environment does not rely on audio or video signals like close circuit television (CCTV) cameras (which are clearly personal data), on the other hand it may give rise to linking, identifiability, detectability, data disclosure, tampering, and spoofing issues [11]. This would result in the processing of data which, by nature or by correlation with other information detained by the MNO, may identify a natural person. Therefore, all sensing targets of Type A, B, and C categories potentially fall into the scope of GDPR application. For this analysis, we assume that personal data (or PII) include, at least, user location and movement patterns, while sensitive data includes, at least, biometric and behavioral data.

*Roles and Responsibilities.* Where the GDPR is applicable, natural and legal persons processing data acquire specific roles and responsibilities defined in legislation. In particular, the natural or legal person, public authority, agency or other body which, alone or jointly with others, determines the purposes and means of the processing of personal data is defined as a *data controller*. The data controller may rely on the work of a data processor which processes personal data on behalf of the controller (*data processor*) and may transfer the data to a third party (*data recipient*). The data controller and, to a certain extent, the data processor are subject to specific obligations vis-à-vis the data subject; the qualification as data controller or data processor is therefore key to determining the lawful behavior of the actor. To stress the peculiarities of the communication systems while still connecting them to the GDPR, we introduce the figure of the *signal controller*, in charge of managing the transmission and reception of sensing-capable signals. This figure is analogous to the data controller defined in the GDPR. However, here the data are physical-layer signals, which have a wider attack surface because they are transmitted through the air. These signals cannot be protected by cryptographic mechanisms alone, but rather require specific solutions, such as *physical-layer security*.

We now associate the stakeholders with the technical features involved in the overall sensing service and the subjective categories of the GDPR, as also depicted in Figure 4:

- *Network end users, belonging to either Type A or C:* They are generally data subjects, as data concerning them is collected and transmitted through the service. However, these end users may become SSCs when they request a 6G Sensing Service. On the other hand, an end user with a ISAC UE providing ISAC information to the network is a SIDP. Accordingly, they can be considered data processors and even data/signal controllers if the processing of data is carried out on their behalf. In the following, we assume, for simplicity, that they are served by their home network, thus no roaming scenario is considered.
- *People not owning a UE, that are, Type B sensing targets*: They are data subjects.
- *MNOs*: They take care of the sensing data collection regarding end users, coordinating their own vendors as well as third-party platforms. They are typically data controllers in most use cases of ISAC. However, they may be required to make data available to the public sector body, e.g. for law enforcement or for exceptional needs under Art. 14 of the EU's *Data Act* [12]; in that case they qualify as data processors.
- *The provider of the radio access network* (RAN) *functionality*: RAN equipment, once typically supplied by vendors, could be provided as a managed service to MNOs. In the latter case, the RAN functionality provider acts as data processors and signal controllers on behalf of the MNO.
- *The provider of the SSP functionality:* SSP functionality vendors may operate sensing data processing on behalf of the MNO, particularly when specialized artificial intelligence/machine learning (AI/ML) and/or public cloud platforms AI/ML or cloud platforms are involved, thus acting as data processors;
- *3-SIDP equipment.* This category includes, e.g., Wi-Fi networks feeding sensing data to the SSP (see Figure 3), thus acting as data controllers and processors.
- *3-SSC platforms*: They can be the MNO itself (especially if transitioning to the so-called *tech-co model*) or an application function (e.g., representing an over-the-top service provider) and can be data recipients. This stakeholder is the final beneficiary of the ISAC information to implement one of the use cases listed above. If they store or further transmit data, they become data processors or even data controllers if the processing of data is carried out on their behalf.
- *Standardization bodies*: RAN and SSP are manufactured (and sold by vendors) in accordance with the 3GPP standard. Therefore, standardization bodies are also stakeholders that must consider the obligations of GDPR.

- *Legal obligations and limits to processing.* The processing of personal data is allowed only based on specific conditions, most notably the consent of the data subject, which must be free, specific, and informed as to the purpose of processing [1, Art. 6 and 7]. Transparency shall be ensured to the data subject before and during the processing. In the telecommunications sector, the ePrivacy Directive introduces even stricter requirements, allowing traffic and location data to be processed only anonymized or with the explicit user's consent.

Data controllers and data processors shall further comply (and be able to show compliance) with the privacy-by-design and privacy-by-default principles, which require to systematically limit data processing to the minimum and embed data protection mechanisms in the structural design of the system, under penalty of considerable administrative fines and to damage liability. The only exception is made for public authorities, which may process personal data without the consent of the data subject to public interest, such as law enforcement or national security, upon conditions established by national law.

In this intricate ecosystem, it is evident that if the extended 6GS (comprising third parties) fails to adhere to privacy-preserving policies, obtain sensing consent, and maintain transparency regarding data subjects, ISAC technology may violate privacy regulations such as GDPR and E-Privacy Directive. This could potentially compromise the security and privacy of individuals' sensing data, resulting in regulatory non-compliance [7]. In particular, the following specific threats may be identified:

- *Unauthorized data collection:* PII and sensitive data may be collected from the TSSA without the knowledge or consent of individuals, thus lacking the mandatory legal basis for lawful processing and transparency.
- *Unauthorized data processing:* Collected sensitive data and PII may be further processed without the knowledge or consent of individuals and/or exploiting technical means that are outside the scope of the contractual agreements with the data subject.
- *Data misuse:* The sensing data collected and processed for one specific purpose could be repurposed for unauthorized applications that involve, e.g., surveillance or profiling, thus lacking the specific consent of the data subject for further processing and transparency.
- *Unawareness of involvement:* The end user may inadvertently participate in sensing data collection and processing, without awareness of the change in roles and without the appropriate safeguards.
- *Unintended disclosure:* High-resolution sensing data and meta information obtained during the sensing process could inadvertently capture identifiable information (e.g., home addresses or personal activities) and expose it to unauthorized parties.

In the light of the kind of data processed, the different roles played by the main actors in the process, and the threats to privacy identified, the use of 6GS appears particularly complex and problematic from a privacy compliance perspective. Table 1 provides an overview of the identified stakeholders, their role as per the GDPR, and the applicable threats.

As already mentioned, when we refer to RAN equipment and SSP functionality vendors, we indirectly refer also to standardization bodies that define the functionalities of the 3GPP standard, after which RANs and SSPs are then designed and implemented. Thus, whatever is relevant for these vendors is also relevant from a definition and design perspective for the manufacturers.

*Table 1: Threat assessment of ISAC-powered mobile networks.*

| Stakeholder | Role | Envisioned threat | Comment |
| --- | --- | --- | --- |
| **Network end users** | Data subject | Unawareness of involvement | Hardware and software of the end user may participate in sensing data collection and processing. |
| | Data recipient | Data misuse | Upon requesting a Sensing Service for a given UE-side application (e.g., a navigation or obstacle-avoidance app), such data are used for a different purpose (e.g., building a behavioral profile of the user for targeted advertising). |
| | Signal controller | Signal misuse | End users may use signals transmitted by the RAN to perform ISAC |
| **People not owning a UE** | Data subject | Unawareness of involvement | Just being present under the coverage of a TSSA yields possible involvement in a sensing service. |
| **MNO** | Data controller | Data misuse | The operator may not frame correctly the conditions of collection and processing of sensing data belonging to the served end users as well as people under the coverage of its network. |
| **Provider of RAN** | Data processor | Unauthorized data collection | The RAN apparatus may fail to collect data from end users and people not owning a UE respecting the conditions set by the operator. |
| | Signal controller | Signal misuse | External attackers may use signals transmitted by the RAN to perform ISAC |
| **Provider of SSP functionality** | Data processor | Unauthorized data processing | The SSP may fail to process collected data from end users and people not owning a UE respecting the conditions set by the operator and/or may leverage external AI/ML tools as well as public cloud platforms to process the collected sensing data. |
| **3-SIDP equipment** | Data controller | Data misuse | The platform may not frame correctly the conditions of collection and processing of sensing data belonging to the sensing targets. |
| | Data processor | Unintended disclosure | Interfacing with SSP may augment the granularity of PII related to the end users and people not owning a UE. |
| **Non-UE SSC equipment** | Data recipient | Data misuse | Upon requesting a Sensing Service for a given NF-side application (e.g., traffic flow monitoring on a highway), such data are used for a different purpose (e.g., tracking the movements of specific individuals across multiple locations). |
| **3-SSC platform** | Data recipient | Data misuse | Upon requesting a Sensing Service for a given external application (e.g., crowd density estimation on a university campus), such data are used for a different purpose (e.g., identifying and profiling individuals attending specific events or gatherings). |

## LAWFUL ISAC SOLUTION FOR CELLULAR NETWORKS

The obligations by GDPR of implementing appropriate technical and organizational measures call for dedicated technical solutions [1, Art. 32], starting from a suitable network architecture able to handle the sensing information and confining its use and diffusion to authorized parties, thus yielding lawful ISAC technology in cellular networks.

*MNO*. It has the main responsibility for sensing data, since it collects them and decides how to handle them. The end user should be entitled to handle sensing data related to him/herself, but, in this case, we must ensure that we do not disclose PII of *other* users without their consent. Examples of policies by MNO that are compatible with the GDPR include:

1. *supervising the entire sensing process*, also in conjunction with third parties outside the 6GS. As part of the agreement with the operator, it holds responsibility for the overall normative compliance of the service it provides to its subscribers.
2. *restricting sensing data collected to specific areas*. We can consider a specific area, e.g., an industrial site, or an academic campus, where users have agreed to give access to be tracked in their movement, for the operations of the industry or for collection of statistics on the campus. In this case, the operator discloses only the sensing information relative to the selective area to the 3-SSC. In this case, the operator can also support the identification of the targets (thus explicitly providing the location and identity of passive users).
3. *tunable retention of sensing data*. Sensing data can be processed immediately after they are acquired and then discarded. For example, sensing acquisitions may be used to detect obstacles and dangerous situations to assist live autonomous driving, but then immediately discarded after use.

*Person with UE*. Here we consider the UE as an SSC, then providing ISAC data to the end user. Although possibly interested in sensing data relative to objects around it, it would be quite problematic to handle such information, even setting a limit on the distance of the sensing target from the UE. Indeed, such sensing information may be relative to other people that are passively tracked, and without their explicit consent, this should be prevented.

*RAN equipment vendor*. Vendors should design and implement technical means comprising, e.g., innovative beamforming techniques capable of preventing unauthorized data extraction while maintaining ISAC functionality and CSI-based user location anonymization. This also may call for a clear distinction between sensing-capable and sensing-resilient waveforms, where the former are deployed exclusively in authorized contexts – a responsibility that falls directly on the signal controller as defined earlier. In addition, the signal controller should be capable of enforcing differentiated service level agreements (SLAs) on sensing performance, ensuring that the resolution and quality of sensing data are tuned to the minimum necessary for each authorized use case – in line with the principle of data minimization.

*SSP functionality vendor*. As the entity responsible for processing sensing data, it should ensure *data processing with proper anonymization procedures*. If a 3-SSC is interested, for example, in obtaining statistical data on the traffic conditions of the main roads covered by the network. In this case, the SSP may pre-process locally the information, ensure that processing is anonymized (e.g., with differential privacy techniques) and then disclose the aggregated anonymized data to the third party. In this case, the sensing data remains unidentifiable.

*3-SIDP platform.* As an authorized external entity providing input sensing data to the 6GS, it must ensure that the data it contributes complies with the same privacy standards required within the network. When acting as a data controller, it must establish a clear legal basis for data collection and ensure transparency towards data subjects. When acting as a data processor, it must operate under a formal data processing agreement with the MNO, limiting data use strictly to agreed purposes and implementing appropriate technical safeguards, such as data minimization and anonymization at the source.

*Non-UE SSC equipment and 3-SSC platforms.* As data recipients, they must ensure that sensing data received from the 6GS is used solely for the declared and agreed purpose. Any further processing or storage must comply with GDPR obligations, including maintaining records of processing activities and implementing appropriate security measures. In particular, 3-SSC platforms operating as over-the-top service providers must establish explicit data sharing agreements with the MNO, clearly defining the scope, retention period, and permitted uses of the received sensing data.

*Standardization bodies*. They play an important role in defining ISAC functions and their interfaces to the rest of the network, in compliance with privacy legislation. These include the design of a) radio signals for ISAC (*data generation and processing*), b) dedicated functions to handle the processing and transferring of the ISAC information within and outside the network (*data control*), c) dedicated control-plane signals to store, manage, and transfer ISAC information. Such interfaces and functionalities enable MNOs and vendors of RAN equipment and UEs to comply with the regulations. An example of such a comprehensive solution is provided in [11].

A summary of the mitigation measures is provided in Table 2.

*Table 2: Overview of the mitigation measures for each identified stakeholder.*

| Stakeholder | Mitigation Approach |
| --- | --- |
| **MNO** | Supervise end-to-end sensing process including third parties; restrict sensing to authorized areas; tunable data retention with immediate discard when possible |
| **Network end users** | Restrict UE-side sensing service requests to declared purposes; prevent exploitation of RAN signals for unauthorized sensing |
| **RAN vendor** | Sensing-resilient vs sensing-capable waveforms; beamforming techniques to prevent unauthorized data extraction; CSI-based user location anonymization; restrict sensing-capable signal transmission to authorized contexts |
| **SSP vendor** | Local pre-processing and anonymization before data leaves the network; differential privacy for aggregated outputs to third parties |
| **3-SIDP platform** | Formal data processing agreements with MNO; source-level anonymization of contributed sensing data; purpose limitation on data provided to SSP |
| **Non-UE SSC equipment** | Purpose limitation enforcement on received sensing data; formal agreements defining scope and retention |
| **3-SSC platform** | Explicit data sharing agreements with MNO; strict purpose limitation; defined retention periods and permitted uses |
| **Standardization bodies** | Define privacy-compliant ISAC radio signal specifications; standardize dedicated functions for ISAC data control; specify control-plane signals for storing, managing and transferring ISAC data; enable privacy-by-design compliance for MNOs and vendors |

## CONCLUSIONS

Adding ISAC capabilities to 6G networks requires a cohesive tech-policy framework that bridges wireless engineering with legal standards. In this paper, we have analyzed the privacy implications of cellular ISAC through the lens of the GDPR, identifying five key privacy threats — unauthorized data collection, unauthorized data processing, data misuse, unawareness of involvement, and unintended disclosure — across all stakeholder categories. A key contribution of our analysis is the introduction of the signal controller concept, which complements the GDPR stakeholder taxonomy by addressing privacy at the physical layer, upstream of data collection. Unlike data controllers and processors, whose responsibilities arise once personal data has been collected, the signal controller represents a preventive first line of defense against unauthorized sensing. Based on our threat assessment, we have outlined mitigation strategies making ISAC *lawful* at both the technical and organizational levels, including physical layer controls, data anonymization, architectural tiering, and MNO policies. Standardization bodies play a particularly important enabling role in this landscape, as embedding privacy-by-design principles into 3GPP and ETSI standards ensures that compliance is a structural feature of the technology rather than an afterthought.

Finally, three open research directions deserve attention: the development of a public trust model [13] for ISAC-enabled networks, the alignment of ISAC capabilities with existing lawful *interception* frameworks — ensuring that sensing information is made available to legal authorities only upon explicit authorized request — and the evolving interaction between ISAC and AI regulation [14, 15], which will require careful monitoring as the legislative landscape matures.